# Physics-informed Deep Diffusion MRI Reconstruction with Synthetic Data: Break Training Data Bottleneck in Artificial Intelligence

Chen Qian[1], Haoyu Zhang[1], Yuncheng Gao[1], Mingyang Han[1], Zi Wang[1], Dan Ruan[1], Yirong Zhou[1], Yu Shen[2], Yiping Wu[2], Chengyan Wang[3], Boyu Jiang[4], Ran Tao[4], Zhigang Wu[5], Jiazheng Wang[5], Liuhong Zhu[6], Yi Guo[6], Taishan Kang[7], Jianzhong Lin[7], Tao Gong[8], Chen Yang[9], Guoqiang Fei[10], Meijin Lin[11], Di Guo[12], Jianjun Zhou[6], Meiyun Wang[2,13], and Xiaobo Qu[1]*

***Abstract*—Diffusion magnetic resonance imaging (MRI) is the only imaging modality for non-invasive movement detection of *in vivo* water molecules, with significant clinical and research applications. Diffusion weighted imaging (DWI) MRI acquired by multi-shot techniques can achieve higher resolution, better signal-to-noise ratio, and lower geometric distortion than single-shot, but suffers from inter-shot motion-induced artifacts. These artifacts cannot be removed prospectively, leading to the absence of artifact-free training labels. Thus, the potential of deep learning in multi-shot DWI reconstruction remains largely untapped. To break the training data bottleneck, here, we propose a Physics-Informed Deep DWI reconstruction method (PIDD) to synthesize high-quality paired training data by leveraging the physical diffusion model (magnitude synthesis) and inter-shot motion-induced phase model (motion phase synthesis). The network is trained only once with 100,000 synthetic samples, achieving encouraging results on multiple realistic in vivo data reconstructions. Advantages over conventional methods include: (a) Better motion artifact suppression and reconstruction stability; (b) Outstanding generalization to multi-scenario reconstructions, including 3 resolutions, 4 b-values, 2 under-samplings, 4 scanners, and 4 centers; (c) Excellent clinical adaptability to patients with verifications by seven experienced doctors (p<0.001). In conclusion, PIDD presents a novel deep learning framework by exploiting the power of MRI physics, providing a cost-effective and explainable way to break the data bottleneck in deep learning medical imaging.**

## I. Introduction

DEEP learning (DL) has made significant achievements in many biomedical magnetic resonance imaging (MRI) tasks [1, 2]. A large amount of paired training data with high-quality labels is commonly required for supervised training [3]. For example, in fast parallel imaging [4], fully sampled k-space data and its under-sampled one are routinely employed as paired labels and inputs [4, 5]. However, high-quality paired data is absent or insufficient in many MRI scenarios [6, 7], resulting in a training data bottleneck in DL image reconstructions. One representative example is the high-resolution diffusion weighted imaging (DWI) MRI.

DWI can detect the *in vivo* water molecule movements non-invasively. It has been widely employed in tumor and stroke diagnosis [8, 9] and tissue microstructure research [10]. Multi-shot interleaved echo planer imaging (ms-iEPI) sequence can obtain DWI with higher resolution, better signal-to-noise ratio, and lower geometric distortion [11-13]. However, during each shot of data acquisition, the subject motion is different. Even slight subject movement on the millimeter scale will cause the significant extra inter-shot phase (motion phase) to be enlarged by strong diffusion gradients [14], resulting in severe image motion artifacts [15, 16]. The physical movement is hardly detectable [13], thus the motion artifacts cannot be removed prospectively. Therefore, the ground truth is absent in multi-shot DWI reconstruction.

Bypassing the problem of data bottleneck but not solving it, current DL methods usually take a lot of time and money to collect a large amount of realistic DWI data and use conventional optimization methods to generate training labels [7, 17, 18]. Many state-of-the-art optimization methods are developed to reconstruct multi-shot DWI, such as MUSE [13], MUSSELS [19, 20], and PAIR [21]. These methods are employed for training label generation and obtain promising results [7, 17]. However, reconstruction of a large number of training labels requires extensive experience and time-

This work was supported in part by the National Natural Science Foundation of China (62331021, 62371410, and 62122064), the Natural Science Foundation of Fujian Province of China under grant (2023J02005), Industry-University Cooperation Projects of the Ministry of Education of China (231107173160805), National Key R&D Program of China (2023YFF0714200), Zhou Yongtang Fund for High Talents Team (0621-Z0332004), the President Fund of Xiamen University (20720220063), and the Xiamen University Nanqiang Outstanding Talents Program.

[1]Department of Electronic Science, Fujian Provincial Key Laboratory of Plasma and Magnetic Resonance, Xiamen University-Neusoft Medical Magnetic Resonance Imaging Joint Research and Development Center, Xiamen University, China. [2]Department of Medical Imaging, Henan Provincial People's Hospital & the People's Hospital of Zhengzhou University, China. [3]Human Phenome Institute, Fudan University, China. [4]United Imaging Healthcare, China. [5]Philips Healthcare, China. [6]Department of Radiology, Zhongshan Hospital, Fudan University (Xiamen Branch), China. [7]Department of Radiology, Zhongshan Hospital Affiliated to Xiamen University, China. [8]Departments of Radiology, Shandong Provincial Hospital Affiliated to Shandong First Medical University, China. [9]Department of Neurosurgery, Zhongshan Hospital, Fudan University (Xiamen Branch), China. [10]Department of Neurology, Zhongshan Hospital, Fudan University, China. [11]Department of Applied Marine Physics and Engineering, Xiamen University, China. [12]School of Computer and Information Engineering, Xiamen University of Technology, China. [13]Biomedical Research Institute, Henan Academy of Sciences. Correspondence should be addressed to Xiaobo Qu (quxiaobo@xmu.edu.cn).

consuming adjustment of parameters of optimization algorithms. Moreover, these algorithms are hard to provide reliable label in some challenging scenarios, such as the non-rigid motion induced high-order phase. PAIR and MUSSELS perform well on synthetic data with the 3-order phase, but failed on the 5-order phase (Figs. 1(m, o)) due to the smoothness assumptions on phases [19-21].

Thus, training with realistic data and reconstructed labels by optimization methods has several drawbacks, hindering the potential power of DL: (a) Labels contain no or inaccurate high-order motion phases; (b) Sizeable realistic database is hard to be acquired due to long acquisition and reconstruction time, limited budget and ethic constraints; (c) Generalization of current DL methods is poor for reconstructions of multi-scenario, such as multi-resolution, multi-b-value, multi-under-sampling, and multi-vendor (Figs. 6-8).

Inspired by an emerging and promising DL paradigm [22, 23] based on the physics-informed synthetic data training, we propose a Physics-Informed Deep Diffusion MRI reconstruction (PIDD). It has two main components (Fig. 2): The physics-informed ms-iEPI DWI data synthesis and a reconstruction network based on motion kernel learning. The preliminary results was presented in a conference paper [24].

The PIDD is a novel framework to overcome the fundamental limitation on high-quality training data in ms-iEPI DWI reconstruction. It is trained only once with 100,000 synthetic samples, achieving encouraging results. Extensive comparison studies with state-of-the-art method shows the advantages of PIDD includes (a) Better motion artifact suppression and reconstruction stability; (b) Outstanding generalization to multi-scenario reconstructions, including multi-resolution, multi-b-value, multi-under-sampling, multi-vendor, and multi-center; (c) Excellent clinical adaptability to patients with verifications by seven experienced doctors ($p<0.001$).

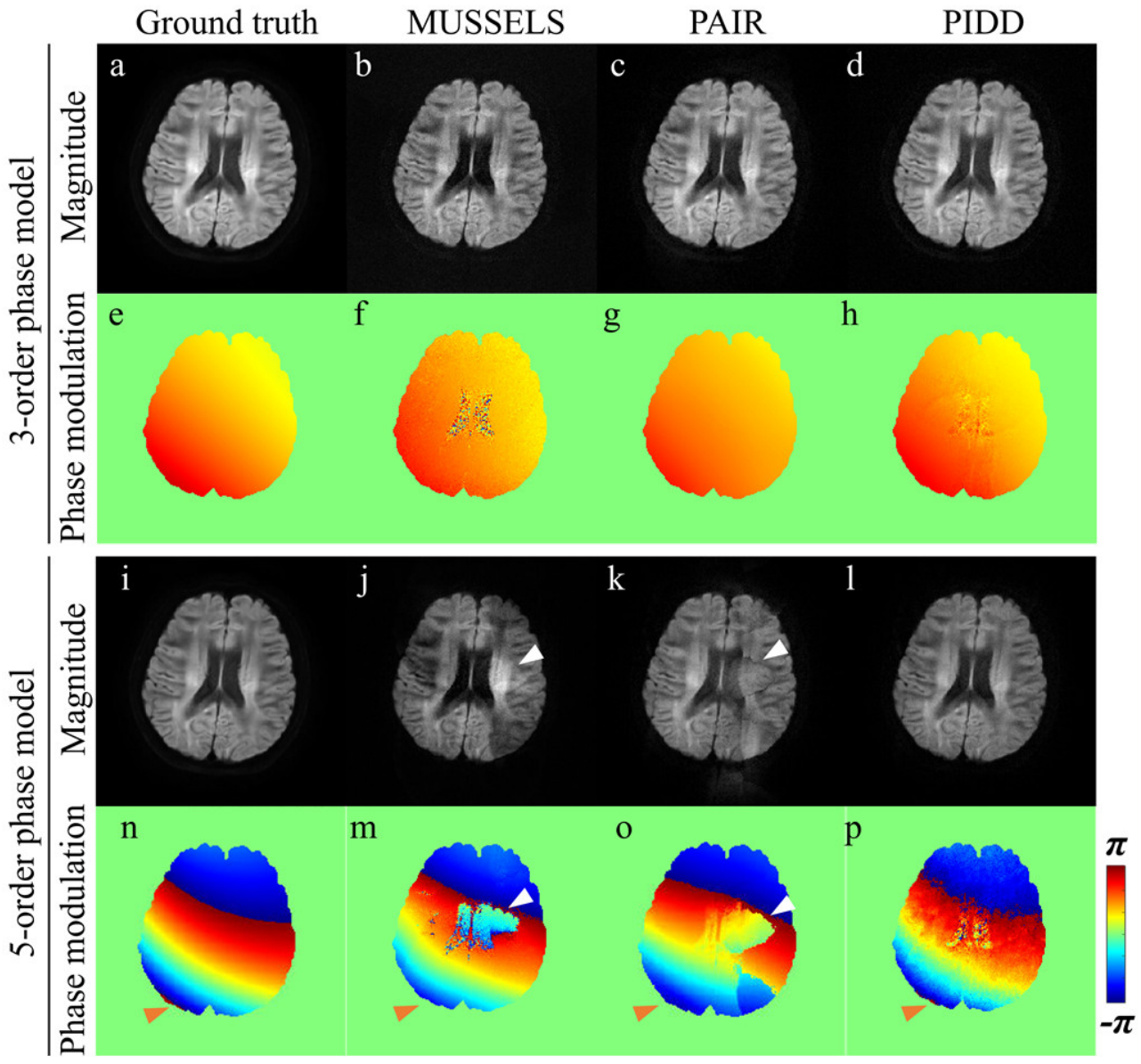

Fig. 1. Comparison of conventional optimization methods (MUSSELS and PAIR) and the proposed PIDD on synthetic data with 3-order and 5-order motion phases. The magnitude and phase modulation between the two shots are shown. PIDD is trained with 100,000 synthetic samples with 5-order motion phases. Incorrect and missing motion phases of MUSSELS and PAIR are marked with white and yellow arrows, respectively.

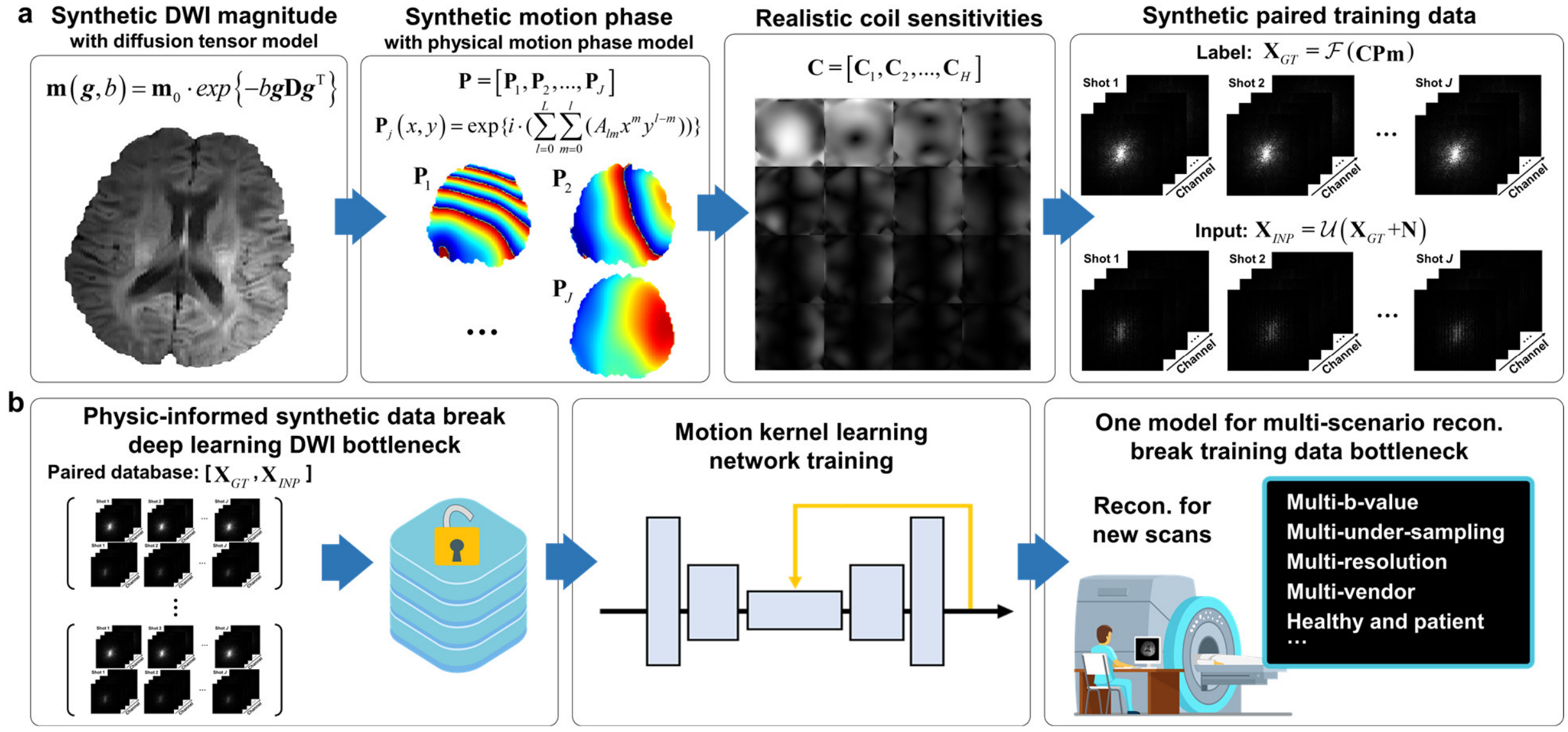

Fig. 2. Physics-informed deep diffusion MRI (PIDD) breaks the bottleneck of obtaining training data in deep learning reconstructions. (a) A large amount of motion phases and DWI magnitudes are synthesized according to physical modeling (diffusion tensro model for DWI magnitude synthesis, and inter-shot motion-induced phase model for motion phase synthesis), which are combined with realistic sensitivity maps to generate training labels. Then, the paired training data are generated from training labels by adding Gaussian noise and undersampling, retrospectively. (b) The proposed motion kernel learning network in PIDD is trained only once with the synthetic database (100,000 samples), having good performance and genralization on multi-scenerio reconstructions, including multi-b-value (1000, 2000, 3000, 4000 s/mm$^2$), multi-resolution (1.5×1.5, 1.2×1.2, 1.0×1.0 mm$^2$), multi-undersampling (partial Fourier at rate of 0.7, 0.8), and multi-vendor (3.0T Philips Ingenia CX, 3.0T Philips Ingenia DNA, and 3.0T United Imaging uMR 890).

## II. PROBLEM FORMULATION

In the DWI acquisition, we assume that the coordinates of any point in the physical coordinate system of the imaging target (also the DWI image coordinate system) are $\vec{r}$ , and the coordinate system of the magnetic resonance gradient field is $\vec{R}$ . Then the position of any point in the imaging target at time *t* in the magnetic resonance gradient field coordinate system is $\vec{R}=\vec{r}+\Delta\vec{R}\left(\vec{r},t\right)$. Thus, the acquired DWI signal of $h^{\text{th}}$ receiver coil can be expressed as [14]:

$$
\begin{aligned}
\mathbf{Y}_h\left(\vec{k}\left(t\right)\right) &= \int \mathbf{C}_h\left(\vec{r}\right)e^{-i2\pi\vec{k}(t)\vec{R}}\mathbf{m}\left(\vec{r}\right)d\vec{r} \\
&= \int \mathbf{C}_h\left(\vec{r}\right)e^{-i2\pi\vec{k}(t)\left(\vec{r}+\Delta\vec{R}(\vec{r},t)\right)}\mathbf{m}\left(\vec{r}\right)d\vec{r} \\
&= \int \mathbf{C}_h\left(\vec{r}\right)e^{-i2\pi\vec{k}(t)\Delta\vec{R}(\vec{r},t)}\mathbf{m}\left(\vec{r}\right)e^{-i2\pi\vec{k}(t)\vec{r}}\,d\vec{r} \\
&= \int \mathbf{C}_h\left(\vec{r}\right)e^{-i\Delta\phi(\vec{r},t)}\mathbf{m}\left(\vec{r}\right)e^{-i2\pi\vec{k}(t)\vec{r}}\,d\vec{r}
\end{aligned}, \quad (1)
$$

where $\mathbf{C}_h\left(\vec{r}\right)$ is the $h^{\text{th}}$ coil sensitivity map, $\mathbf{m}\left(\vec{r}\right)$ is the transverse magnetization with diffusion information, and $\vec{k}\left(t\right)$ is the coordinates in the k-space. Thus, the extra motion phase in Eq. (1) caused by the displacement $\Delta\vec{R}\left(\vec{r},t\right)$ is:

$$
\Delta\phi\left(\vec{r},t\right)=2\pi\cdot\vec{k}\left(t\right)\cdot\Delta\vec{R}\left(\vec{r},t\right)=\gamma\int_0^t\Delta\vec{R}\left(\vec{r},t'\right)\cdot\vec{G}\left(t'\right)dt', \quad (2)
$$

where $\vec{G}$ is vector for gradient field. Assuming that the duration of the gradient field is *T*, the motion phase will be:

$$
\Delta\phi\left(\vec{r}\right)=\gamma\int_0^T\Delta\vec{R}\left(\vec{r},t\right)\cdot\vec{G}\left(t\right)dt, \quad (3)
$$

where motion displacement $\Delta\vec{R}$ determines the characteristics of the motion phase $\Delta\phi$ . Under different displacement conditions, the motion phase takes the following forms:

$$
\Delta\phi\left(\vec{r}\right)=\begin{cases}
0, & \Delta\vec{R}\left(\vec{r},t\right)=0, \\
\Delta R\cdot\phi_G, & \Delta\vec{R}\left(\vec{r},t\right)=\Delta R, \\
\theta\cdot\vec{r}\cdot\phi_G, & \Delta\vec{R}\left(\vec{r},t\right)=\theta\cdot\vec{r}, \\
\sum_{l=0}^{L}\alpha_l\vec{r}^{\,l}\cdot\phi_G, & \Delta\vec{R}\left(\vec{r},t\right)=\sum_{l=0}^{L}\alpha_l\vec{r}^{\,l},
\end{cases} \quad (4)
$$

where $\phi_G=\gamma\int_0^T\vec{G}\left(t\right)dt$ is a constant. The rigid body translational ( $\Delta R$ is a constant independent of position $\vec{r}$ ) and rotation ( $\theta$ is the rotation degree) will introduce 0 and 1 order motion phase about $\vec{r}$ , respectively; when non-rigid body motion occurs, the motion displacement becomes more complex and becomes a higher-order function of $\vec{r}$ (*L* represents the highest order), the motion phase also becomes a *L*-order polynomial function of $\vec{r}$ .

The above analysis on the characteristics of motion phase inspires us to use *L*-order polynomial functions to approximate, fit, and generate the measured motion phase (Fig. 4).

In the ms-iEPI DWI acquisition, the discretized form of k-space signal of $j^{\text{th}}$ shot in Eq. (1) donates:

$$
\mathbf{Y}_{h,j}\left(kx,ky\right)=\int\mathbf{C}_h\left(x,y\right)e^{-i\left(\Delta\phi_j(x,y)\right)}e^{-i2\pi(kx,ky)\cdot(x,y)}\mathbf{m}\left(x,y\right)dxdy, \quad (5)
$$

where $(x, y)$ and $(kx, ky)$ are the two-dimensional coordinates of the image domain and k-space respectively. After inverse Fourier transform, the image domain signal collected by the $h^{\text{th}}$ channel is:

$$
\mathbf{I}_{h,j}=\mathbf{C}_h e^{-i\left(\Delta\phi_j\right)}\mathbf{m}=\mathbf{C}_h\mathbf{P}_j\mathbf{m}, \quad (6)
$$

where $\mathbf{P}_j=e^{-i\left(\Delta\phi_j\right)}$ is the image phase of the $j^{\text{th}}$ shot (background phase is ignored here).

Since each shot only uniformly samples the partial k-space of a DWI image, the k-space signal of $j^{\text{th}}$ shot of $h^{\text{th}}$ channel is

$$
\mathbf{Y}_{h,j}=\mathcal{U}_j\mathcal{F}\mathbf{C}_h\mathbf{P}_j\mathbf{m}+\mathbf{N}, \quad (7)
$$

where $\mathcal{F}$ is the Fourier transform, $\mathcal{U}_j$ is the filling trajectory of the $j^{\text{th}}$ shot data in the k-space, namely the sampling mask, $\mathbf{N}$ is the Gaussian noise.

Therefore, the core problem of ms-iEPI DWI reconstruction is to reconstruct the phase $\mathbf{P}_j$ and magnitude $\mathbf{m}$ from the sampled multi-shot multi-channel k-space data $\mathbf{Y}$.

## III. PROPOSED METHOD

### *A. Physics-informed training data synthesis in PIDD*

In this part, we first introduce the training data synthesis in PIDD. The whole process includes: (1) Synthesize DWI magnitude images $\mathbf{m}$ with specific b-values and diffusion directions according to a physical diffusion model; (2) Synthesize motion phase $\mathbf{P}_j$ with an analytically polynomial phase model; (3) Multiply magnitude images with synthetic motion phases to get multi-shot images; (4) Multiply each shot of image with a realistic coil sensitivity $\mathbf{C}$; (5) Transform each shot of image of each channel into k-space as the ground truth (label) for network training; (6) Generate network inputs by under-sampling the noisy k-space according to the sampling patterns of ms-iEPI sequence.

In step (4), 300 realistic coil sensitivity maps are estimated by ESPIRIT [25] from 16-channel $T_2$ weighted MRI data from the Fast MRI database [26].

In step (5), the synthesized ground truth k-space (label) is:

$$
\mathbf{X}_{GT}=\mathcal{F}\left(\mathbf{CPm}\right), \quad (8)
$$

where $\mathcal{F}$ is the Fourier transform, $\mathbf{P}$ = [$\mathbf{P}_1$; $\mathbf{P}_2$; …, $\mathbf{P}_J$] is the motion phases.

In step (6), the noisy k-space is obtained by adding Gaussian noise $\mathbf{N}$ to ground truth k-space to decrease its signal-to-noise-ratio to the range of 10-50 dB. Then, the input data is:

$$
\mathbf{X}_{INP}=\mathcal{U}\mathcal{F}\left(\mathbf{CPm}\right)+\mathbf{N}, \quad (9)
$$

where $\mathcal{U}$ is the sampling patterns of ms-iEPI sequence.

By randomly setting physics generation parameters (b-value, diffusion direction, polynomial coefficients, and noise level) within a given range and repeating the above steps, we generate a total of 100,000 pairs of training data with ground truth labels.

The following parts describe the detailed synthesis of DWI magnitude $\mathbf{m}$ and motion phases $\mathbf{P}_j$ in steps (1) and (2).

#### *1) Physics-informed DWI magnitude synthesis*

The DWI magnitude is generated according to a physical diffusion model [27]. Here, we choose the diffusion tensor imaging (DTI) model [28]:

$$
\mathbf{m}\left(\boldsymbol{g},b\right)=\mathbf{m}_0\cdot exp\left\{-b\mathbf{g}\mathbf{D}\mathbf{g}^{\mathrm{T}}\right\}, \quad (10)
$$

where $\mathbf{m}(\boldsymbol{g}, b)$ is the magnitude of the diffusion signal with diffusion direction $\boldsymbol{g}$ and b-value $b$, and $\mathbf{m}_0$ is a non-diffusion image (b-value=0). The $\mathbf{D}$ is the diffusion tensors. The superscript T is the transpose. Once the high-quality diffusion tensors $\mathbf{D}$ and $\mathbf{m}_0$ are known, and imaging experiment parameters ($\boldsymbol{g}$ and $b$) are provided, the magnitude of diffusion signal $\mathbf{m}$ can be synthesized according to Eq. (10).

We acquire a set of DWI images from a 3.0T Philips Ingenia CX scanner to get a high-quality diffusion tensor. The image set contains two non-diffusion images and 64 DWI images (b-value=1000 s/mm$^2$) with uniformly distributed diffusion directions. Then, the diffusion tensor is calculated with the Python package DIPY [29].

When the high-quality diffusion tensor and non-diffusion image are known in Eq. (1), the magnitude of diffusion signal $\mathbf{m}(\boldsymbol{g}, b)$ with given diffusion direction and b-value can be synthesized (Fig. 3). The given $\boldsymbol{g}$ are [0,0,1], [0,1,0], [1,0,0], and $b$ are 1000, 2000, 3000, and 4000 s/mm$^2$.

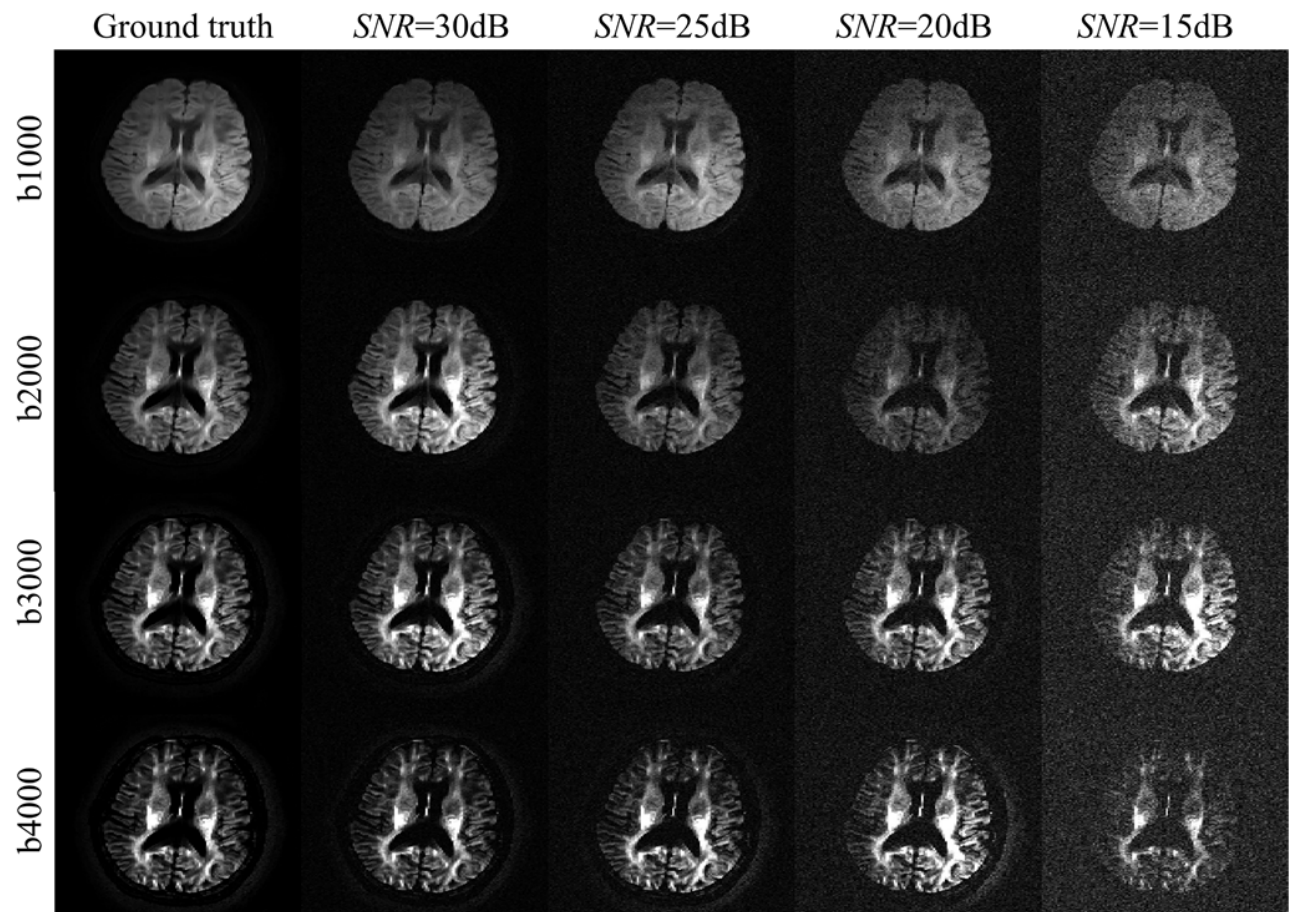

**Fig. 3.** Representatively synthesized DWI magnitude images with the added noise. The diffusion direction is [0,0,1], and b-values are 1000, 2000, 3000, and 4000 s/mm$^2$. The Gaussian noise is added in the k-space to decrease the SNR to 30, 25, 20, and 15 dB, respectively.

### *2) Physics-informed motion phase synthesis*

The motion phase is generated according to a physically derived polynomial phase model [14] with increased orders:

$$\mathbf{P}_j(x,y) = e^{i\cdot\Delta\phi_j(x,y)} = \exp\{i\cdot(\sum_{l=0}^{L}\sum_{m=0}^{l}(A_{lm}x^m y^{l-m}))\}, \quad (11)$$

where $(x, y)$ is the image coordinate, $L$ is the order of the polynomial model, and $A^{lm}$ is the coefficients of $x^m y^{l-m}$. Once these parameters are set, a large amount of motion phases can be generated by Eq. (11).

The major sources of macroscopic motion in brain DWI are involuntary head motion and pulsatile brain motion [14, 30]. The former can be modeled as rigid motion, such as shift and rotation, introducing low-order motion phases. The latter is generally considered brain parenchymal motions and cerebrospinal fluid pulsations [30, 31], resulting in high-order motion phases. Thus, we use a high-order polynomial motion phase model to synthesize the motion phase.

Experimentally, this phase model is designed to fit the realistic motion phase well. We obtain realistic motion phases by the conventional optimization method PAIR [21], and test the performance of phase models of different orders on fitting the real phase (Fig. 4). A larger $L$ shows a better fitting of complex shot phases. In our scheme, $L = 5$ is selected to balance computational complexity and accuracy. For $l = 0, 1, 2, 3, 4, 5$, $A^{lm}$ are randomly distributed in [-π, π), [-π, π), [-π, π), [-π/2, π/2), [-π/2, π/2), [-π/2, π/2).

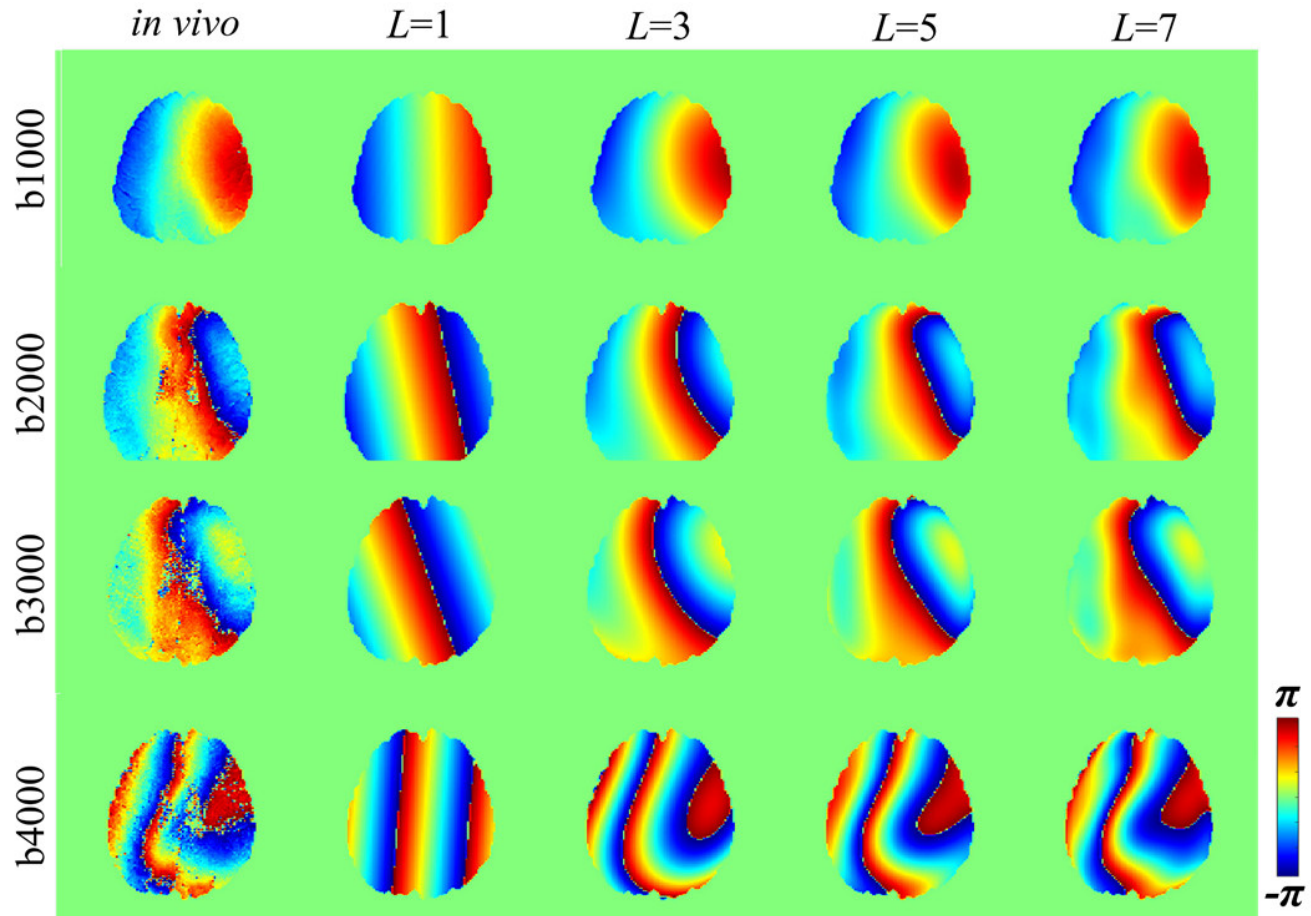

**Fig. 4.** Representatively synthesized motion phases. The in vivo motion phases of 4-shot DWI with different b-values (1000, 2000, 3000 s/mm$^2$) and corresponding synthetic motion phases with different polynomial order ($L$ = 1, 3, 5, 7) are shown.

## *B. Motion kernel learning network in PIDD*

In this part, we introduce another part of PIDD, namely the reconstruction network based on motion kernel learning.

### *1) Motion kernel interpolation between shots*

In the ms-iEPI DWI, the $i^{th}$ shot image $\mathbf{I}_i$ and the $j^{th}$ shot image $\mathbf{I}_j$ have the relationship:

$$\mathbf{I}_i = \mathbf{P}_i\mathbf{m} = \mathbf{P}_i\frac{\mathbf{P}_j^\dagger\mathbf{P}_j}{\mathbf{P}_j^\dagger\mathbf{P}_j}\mathbf{m} = \mathbf{P}_i\frac{\mathbf{P}_j^\dagger}{\mathbf{P}_j^\dagger\mathbf{P}_j}\mathbf{P}_j\mathbf{m} = \mathbf{P}_i\frac{\mathbf{P}_j^\dagger}{\mathbf{P}_j^\dagger\mathbf{P}_j}\mathbf{I}_j = \hat{\mathbf{P}}_{ij}\mathbf{I}_j, \quad (12)$$

where $\mathbf{P}_i\frac{\mathbf{P}_j^\dagger}{\mathbf{P}_j^\dagger\mathbf{P}_j} = \hat{\mathbf{P}}_{ij}$ is the motion phase modulation, and the superscript $\dagger$ is the complex conjugate.

Performing the Fourier transform on the two side of Eq. (1), and we get a convolution form in k-space:

$$\mathbf{X}_i = \mathcal{G}_{ij}\mathbf{X}_j, \quad (13)$$

where $\mathbf{X}_i$ and $\mathbf{X}_j$ are the k-space of $i^{th}$ and $j^{th}$ shot images and $\mathcal{G}_{ij}$ is a convolution kernel determined by $\hat{\mathbf{P}}_{ij}$. Furthermore, we build a structured matrix $\mathbf{Q}$ consisting of motion phase modulations $\hat{\mathbf{P}}_{ij}$, which has the relationships between all shot images:

$$\mathbf{QI} = \frac{1}{J-1}\begin{bmatrix} 0 & \hat{\mathbf{P}}_{12} & \hat{\mathbf{P}}_{13} & \cdots & \hat{\mathbf{P}}_{1J} \\ \hat{\mathbf{P}}_{21} & 0 & \hat{\mathbf{P}}_{23} & \cdots & \hat{\mathbf{P}}_{2J} \\ \hat{\mathbf{P}}_{31} & \hat{\mathbf{P}}_{32} & 0 & \cdots & \hat{\mathbf{P}}_{3J} \\ \vdots & \vdots & \vdots & 0 & \vdots \\ \hat{\mathbf{P}}_{J1} & \hat{\mathbf{P}}_{J2} & \hat{\mathbf{P}}_{J3} & \cdots & 0 \end{bmatrix}\begin{bmatrix} \mathbf{I}_1 \\ \vdots \\ \mathbf{I}_j \\ \vdots \\ \mathbf{I}_J \end{bmatrix} = \begin{bmatrix} \mathbf{I}_1 \\ \vdots \\ \mathbf{I}_j \\ \vdots \\ \mathbf{I}_J \end{bmatrix}. \quad (14)$$

Perform the Fourier transform on Eq. (14) by substituting multiplication in the image space with the convolution operator in Eq. (13), and we get:

$$\mathcal{G}\mathbf{X} = \frac{1}{J-1}\begin{bmatrix} 0 & \mathcal{G}_{12} & \mathcal{G}_{13} & \cdots & \mathcal{G}_{1J} \\ \mathcal{G}_{21} & 0 & \mathcal{G}_{23} & \cdots & \mathcal{G}_{2J} \\ \mathcal{G}_{31} & \mathcal{G}_{32} & 0 & \cdots & \mathcal{G}_{3J} \\ \vdots & \vdots & \vdots & 0 & \vdots \\ \mathcal{G}_{J1} & \mathcal{G}_{J2} & \mathcal{G}_{J3} & \cdots & 0 \end{bmatrix}\begin{bmatrix} \mathbf{X}_1 \\ \vdots \\ \mathbf{X}_j \\ \vdots \\ \mathbf{X}_J \end{bmatrix} = \begin{bmatrix} \mathbf{X}_1 \\ \vdots \\ \mathbf{X}_j \\ \vdots \\ \mathbf{X}_J \end{bmatrix} = \mathbf{X}, \quad (15)$$

where $\mathcal{G}$ is a set of interpolation kernels between the shot data in the k-space. Then, we proposed a k-space interpolation model for DWI reconstruction with motion kernel:

$$\min_{\mathbf{X}} \sum_{h}^{H} \left\| \mathbf{Y}_h - \mathcal{U}\mathcal{F}\mathbf{C}_h\mathcal{F}^{-1}\mathbf{X} \right\|_{\mathrm{F}}^2 + \frac{\lambda}{2}\left\| \mathcal{G}\mathbf{X} - \mathbf{X} \right\|_{\mathrm{F}}^2, \quad (16)$$

where $\mathbf{Y}_h$ is the sampled $h^{\text{th}}$ channel k-space data, $H$ is the number of channels, $\mathcal{F}^{-1}$ is the inverse Fourier transform operator, $\lambda$ is the regularization parameter, and $\|\cdot\|_{\mathrm{F}}^2$ presents the squared Frobenius norm.

We take the Projection Onto Convex Sets algorithm to solve this model, and the $k^{\text{th}}$ iteration contains:

$$\begin{cases} \mathbf{G}_h^k = \mathbf{C}_h\mathcal{F}^{-1}\mathbf{X}^{k-1} + \lambda\mathcal{F}^{-1}\mathcal{U}^*\left(\mathbf{Y}_h - \mathcal{U}\mathcal{F}\mathbf{C}_h\mathcal{F}^{-1}\mathbf{X}^{k-1}\right) \\ \mathbf{Z}^k = \mathcal{F}\left(\sum_h^H \mathbf{C}_h^*\mathbf{G}_h^k\right) \\ \mathbf{X}^{k+1} = \mathcal{G}\left(\mathbf{Z}^k\right) \end{cases}. \quad (17)$$

To make it work in real scenarios where $\mathcal{G}$ is unknown, we replace $\mathcal{G}$ with a learnable network $\mathcal{N}$ in the following motion kernel learning.

### 2) Motion kernel learning

By unrolling the solving process in Eq. (17), we obtain the architecture of the motion kernel learning network (Fig. 5). It has $K = 10$ blocks, and the $k^{\text{th}}$ block contains three modules.

**Data consistency module:** The projection operator for data consistency is used to project the initial image onto a convex set, which contains all images that have the same value as the acquired data at sampled k-space positions:

$$\mathbf{G}_h^k = \mathbf{C}_h\mathcal{F}^{-1}\mathbf{X}^{k-1} + \lambda\mathcal{F}^{-1}\mathcal{U}^*\left(\mathbf{Y}_h - \mathcal{U}\mathcal{F}\mathbf{C}_h\mathcal{F}^{-1}\mathbf{X}^{k-1}\right) \quad (18)$$

$$\mathbf{Z}^k = \mathcal{F}\left(\sum_h^H \mathbf{C}_h^*\mathbf{G}_h^k\right), \quad (19)$$

where the superscript * is the adjoint operation, and the regularization parameter $\lambda$ is set to 1. This module has no learnable parameters.

**Motion kernel learning module:** a convolution network $\mathcal{N}$ is used as the projection operator to learn the motion kernels:

$$\mathbf{X}^{k+1} = \mathcal{N}\left(\mathbf{Z}^k\right), \quad (20)$$

In each block, $\mathcal{N}$ has 6 layers, and each is composed of 48 convolution kernels of size 3. The parameters can be updated in the supervised training stage. In the reconstruction stage, these convolution kernels in $\mathcal{N}$ play the role of motion kernels $\mathcal{G}$ to interpolate missing k-space data among the shots.

**Output module:** k-space conjugate symmetry property is employed to extrapolate missing k-space data for image detail preservations in partial Fourier under-sampling reconstructions as follows:

$$\mathbf{X}^{out} = \mathcal{P}^*\left(\mathrm{SVT}_\varepsilon\left(\mathcal{P}\mathbf{Z}^{K+1}\right)\right), \quad (21)$$

where $\mathrm{SVT}_\varepsilon$ is the singular value thresholding operator [32], and the first $\varepsilon$ singular values are saved. $\mathcal{P}$ and $\mathcal{P}^*$ are the operator and adjoint operator for the structured low-rank matrix construction in [21, 33].

In the training stage, the loss function is:

$$\mathcal{L} = \frac{1}{KS}\sum_{s=1}^{S}\sum_{k=1}^{K}\left\| \mathbf{X}^{k,s} - \mathbf{X}_{GT}^{s} \right\|_F^2, \quad (22)$$

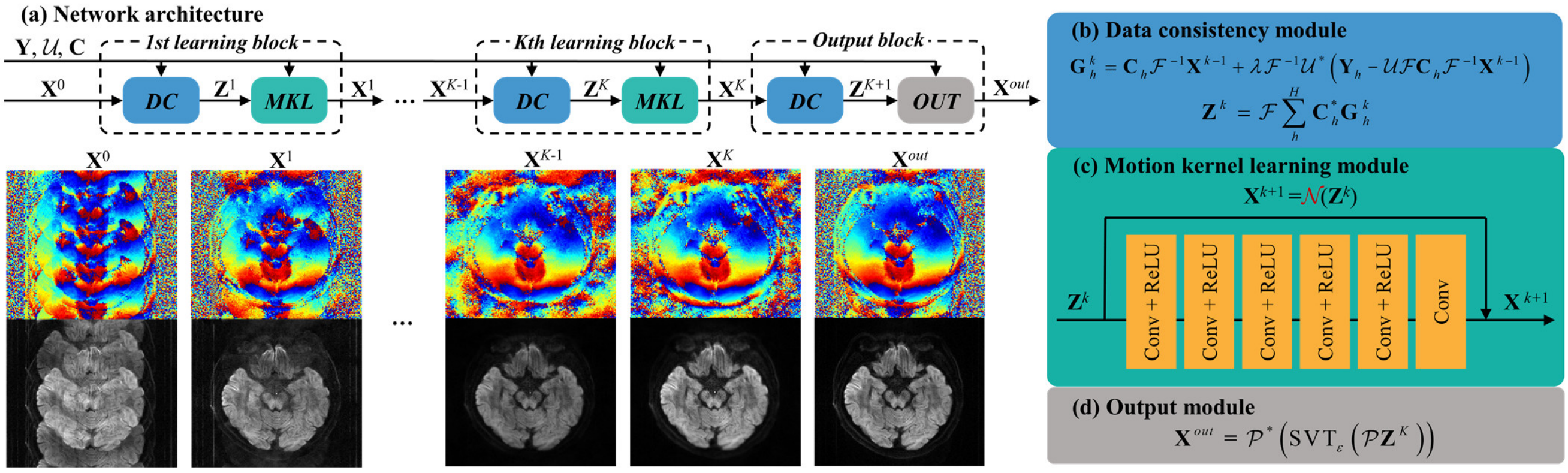

Fig. 5. The architecture of the motion kernel learning network in PIDD. (a) The motion artifacts is removed step by step through the iterative blocks. Each block includes (b) a data consistency (DC) module, (c) a motion kernel learning (MKL) module, and (d) an output (OUT) module. Note: Conv is the convolutional layer, and ReLU is the activation function. $\mathbf{Y}$ is the sampled multi-shot multi-channel k-space data, $\mathbf{C}$ is the coil sensitivity maps, $\mathcal{U}$ is the sampling mask, and $\mathbf{X}^0 = \mathcal{F}^{-1}\mathcal{U}^*\mathbf{Y}$ is the k-space of initial DWI image.

where $S$ is the number of training samples, $K$ is the output of the $k^{th}$ block with $s^{th}$ training samples as input, $\mathbf{X}_{GT}^{s}$ is the $s^{th}$ label, and $\mathbf{X}^{k,s}$ is the output of $k^{th}$ block.

The weights of the reconstruction network are Xavier initialized and trained with $S$ = 100,000 synthetic training samples by Adam optimizer. Its initial learning rate is 0.001, with an exponential decay of 0.99. The batch size is 1, and the training takes about 120 hours on a Nvidia Tesla T4 GPU (16 GB memory) with TensorFlow 1.15.0.

## IV. Experimental Results

### A. Datasets

The PIDD trained with 100,000 synthetic data is compared with state-of-the-art optimization and deep learning methods on four realistic 4-shot iEPI DWI datasets. All realistic DWI databases are from our collections, and all experiments are Institutional Review Board-approved and volunteer-informed. The detailed acquisition specifications for the datasets are provided below.

**DATASET I (Healthy volunteer DTI):** Acquired on a 3.0T scanner in Xiamen, China (Ingenia CX, Philips Healthcare). Seven healthy volunteers are included. Matrix size is 180 × 180 (1.5 × 1.5 mm$^2$), b-values are 0 and 1000 s/mm$^2$, and diffusion direction is 12.

**DATASET II (Healthy volunteer DWI):** Acquired on a 3.0T scanner in Shanghai, China (uMR 890, United Imaging Healthcare). One healthy volunteer is scanned. Matrix sizes are 160 × 160 (1.5 × 1.5 mm$^2$) and 230 × 224 (1.0 × 1.0 mm$^2$), and b-values are 0, 1000, 2000, 3000, and 4000 s/mm$^2$.

**DATASET III (Patient DTI):** Acquired on a 3.0T scanner in Zhengzhou, China (Ingenia DNA, Philips Healthcare). Nine patients are included. Matrix size is 220 × 180 (1.0 × 1.2 mm$^2$), b-values are 0 and 1000 s/mm$^2$, diffusion directions are 12, and partial Fourier under-sampling rate is 0.6. Other clinical protocols are also acquired as reconstruction references, including T2 weighted, contrast enhanced T1 weighted, fluid attenuated inversion recovery, and single-shot EPI DWI images.

### B. Experiments settings

For comparative study, two optimization methods MUSSELS [19, 20] and PAIR [21], and a state-of-the-art deep learning method MODL [7] are used. MUSSELS is one of most employed ms-iEPI DWI reconstruction method, which exploits cross-shot annihilating filter relations deduced from smooth phase modulations, and the interpolates missing data in k-space from multi-shot data by a structured low-rank matrix completion formulation. MoDL is an unrolled deep learning network inspired by the MUSSELS reconstruction. With MUSSELS results as training labels, MoDL can obtain close performance as MUSSELS but much faster reconstruction speed. PAIR employs an iteratively joint estimation model with the paired priors of phase and magnitude to regularize the reconstruction.

In this work, the proposed PIDD is trained with 100,000 synthetic data. The MoDL is trained with 1296 realistic samples (6 cases from **DATASET I**, 108 slices × 12 directions = 1296 training samples), and training labels are reconstructed by PAIR. The one synthetic data trained PIDD and one *in-vivo* healthy data trained MoDL model are used for all four realistic datasets reconstructions, including matched and mismatched scenarios.

All methods are running on a server equipped with Intel Xeon Silver 4210 CPUs (256 GB RAM), and Nvidia Tesla T4 GPU (16 GB memory). PIDD and MoDL are implemented with TensorFlow 1.15.0, while the PAIR and MUSSELS are implemented with MATLAB. The codes of all comparison methods are shared by their authors.

Before reconstructions, the echo-planar imaging ghost is carefully corrected by a widely employed reference-free method [34]. Few ghost residuals, however, may still exist in some images. All DWI data are coil compressed [35] to no more than 16 channels to reduce computational complexity.

For reconstructions, coil sensitivity maps are estimated by ESPIRIT [25] with non-diffusion data (b-value = 0). DTI metrics such as FA, MD, and diffusion tensors are calculated using the Python package DIPY [29].

In all experiments, the reconstructed multi-shot images are finally displayed after combining all shot images by the square root of sum of squares.

Three objective metrics are adopted to assess reconstructions. ghost-to-signal ratio (GSR) [36] is utilized to evaluate the motion artifact suppression quantitatively, and a lower GSR means better artifact removal. Relative $l_2$ norm error (RLNE) and peak signal-to-noise ratio (PSNR) are employed for evaluation of image details and noise. The lower RLNE and higher PSNR indicate better image details recovery and noise suppression.

Three clinical-concerned subjective metrics are adopted in the reader study, including image signal-to-noise ratio (SNR), artifact suppression, and overall image quality. The single-shot DWI and DTI images are provided as references. Each criterion's score ranges from 0 to 5 with precision of 0.1 (i.e., 0~1: Non-diagnostic; 1~2: Poor; 2~3: Adequate; 3~4: Good; 4~5: Excellent). The reader study is performed on CloudBrain-ReconAI [37], which is free to access at https://csrc.xmu.edu.cn/CloudBrain.html.

### C. Comparison study on matched reconstruction

In this section, PIDD is compared with two optimization methods, including PAIR [21] and MUSSELS [19], and one DL method MoDL [7] on 144 testing samples from **DATASET I** (Figs. 6). MoDL is trained with realistic data (six cases from **DATASET I**, 108 slices × 12 directions = 1296 training samples), and the training labels is reconstructed by PAIR with optimized parameters. PIDD is trained once with 100,000 synthetic samples.

Serious motion artifacts (Figs. 6(a, f)) are observed if a trivial inverse Fourier transform (IFT) is applied, resulting in high GSR and many outliers (Figs. 6(k)). This observation implies that motion-induced phase and artifacts have a significant variance among different data. Thus, robust reconstruction of 144 testing samples is challenging.

State-of-the-art comparison methods are sensitive to samples. Algorithm parameters of PAIR and MUSSELS are optimized to successfully remove most artifacts for the first sample (Figs. 6(b, c)) but may degrade image quality for another sample that has a different slice and diffusion direction (Figs. 6(g, h)). This sensitivity also limits the performance of MoDL since its training labels are PAIR reconstructions. Thus, the GSR of MoDL is slightly worse than that of PAIR (Figs. 6(k)). Even though, MoDL is still very valuable as its reconstruction time is greatly reduced than PAIR (Figs. 6(l)).

PIDD synthetizes the training data much faster (more than 100 times in Figs. 6(l)) than realistic labels generation and the reconstruction performance is more robust than all compared methods. It suppresses the motion artifacts very well on these samples (Figs. 6(e, j)) and achieves the lowest GSR (Figs. 6(k)).

We further compare their performance on the retrospective partial Fourier under-sampling (under-sampling rate 0.7, and 0.8). The best GSR, PSNR, and RLNE show that, PIDD removes artifacts and preserves image details better than other methods under partial Fourier under-sampling (TABLE I)

TABLE I

GSR / PSNR / RLNE OF PARTIAL FOURIER RECONSTRUCTIONS (MEAN ± STD)

| Under-sampling | Method | GSR ($\times 10^{-2}$) | PSNR (dB) | RLNE ($\times 10^{-2}$) |
|---|---|---|---|---|
| 0.8 Partial Fourier | IFT | 59.37±32.35 | / | / |
| | PAIR | 6.15±3.38 | 37.33±3.06 | 5.76±1.82 |
| | MUSSELS | 24.17±17.14 | 37.33±2.72 | 5.86±0.89 |
| | MoDL | 11.25±6.09 | 37.92±2.10 | 5.34±0.62 |
| | PIDD | **2.55±1.96** | **39.92±2.71** | **4.11±0.45** |
| 0.7 Partial Fourier | IFT | 60.36±33.33 | / | / |
| | PAIR | 5.96±3.34 | 35.50±2.95 | 7.07±1.61 |
| | MUSSELS | 23.43±16.67 | 35.35±2.73 | 7.36±1.10 |
| | MoDL | 11.67±6.65 | 35.31±2.13 | 7.29±0.84 |
| | PIDD | **2.55±1.97** | **37.29±2.79** | **5.56±0.58** |

Note: The retrospective partial Fourier under-sampling on fully-sampled test data from **DATASET I** is used. The means and standard deviations are computed over all 144 reconstructed results, respectively. When calculating PSNR and RLNE, the reconstructed results of fully-sampled data by each method are used as corresponding references

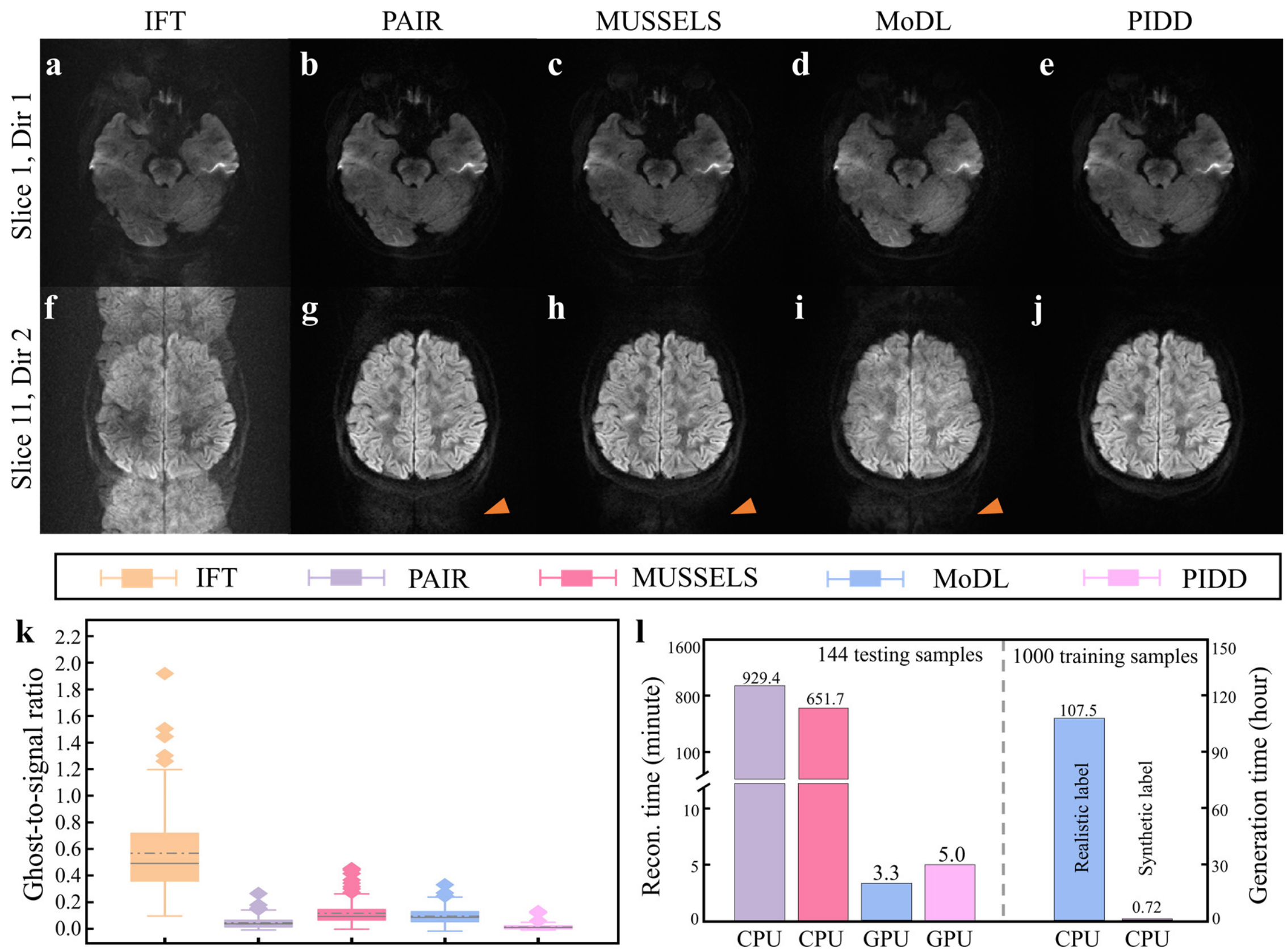

**Fig. 6.** Comparison study on the **DATASET I** of healthy volunteer DTI. (a)-(j) are the reconstructed DWI images by the methods for comparisons and PIDD. Motion artifact residuals are marked by yellow arrows (the second row). (k) is the ghost-to-noise ratio (GSR) calculated on 144 DWI images. (l) are the comparisons of reconstruction time of PAIR, MUSSELS, MoDL, and PIDD, respectively, and label generation time of MoDL, and PAIR. Note: IFT is the inverse Fourier transform. The hardware platform employed for all algorithms is a server with Intel Xeon Silver 4210 CPUs (256 GB RAM), and Nvidia Tesla T4 GPU (16 GB memory).

### *D. Mismatched comparison reconstructions in multiple generalizable scenarios*

The generalization ability of PIDD is tested on **DATASET II** (Fig. 7), which contains DWI data with multi-b-value and multi-resolution. Reconstruction parameters of PAIR are optimized on each image to get the best motion artifact removal. Thus, PAIR results are used as reference to evaluate the performance of two DL methods, MoDL and PIDD.

For MoDL (Fig. 7(h-m)), the large mismatch between the training (**DATASET I**) and testing data (**DATASET II**) results in noticeable artifacts. Its trace DWI images combined from three diffusion directions have severe mistakes (arrows in Figs. 7(k, m)), including structural errors, incorrect contrast, and significant noise corruption with ultra-high b-values (b-value=4000 s/mm$^2$).

For PIDD (Figs. 7(o-u)), it shows great generalization on multi-b-value (1000, 2000, 3000, and 4000 s/mm$^2$), and multi-resolution (1.5×1.5 mm$^2$ and 1.0×1.0 mm$^2$).

This good generalization of PIDD is due to the flexible training data synthesis. First, multi-b-value DWI images can be easily synthesized by the diffusion tensor model. Second, the Gaussian noise added to k-space has a wide range to cover diverse noise levels in realistic multi-b-value data. Last but not least, the high-order polynomial model can cover the non-rigid motion induced high-order phase.

These results demonstrate that PIDD achieves a better generalization over the compared DL method under multiple DWI reconstruction scenarios.

### *E. Adaptability to patient data*

This section examines the adaptability of PIDD on patient DTI data from **DATASET III**, which is essential in clinical diagnosis. PAIR, MoDL, and PIDD reconstruct DTI data collected from 9 patients, and each patient's DTI has 18 slices and 16 diffusion directions (2592 samples in total). MoDL is still trained with six cases from **DATASET I**. PIDD is trained once with 100,000 synthetic samples. Thus, both MoDL and PIDD have never seen any patient data.

MRI images of a 25-year-old female patient after surgery for brain diffuse astrocytoma are shown in Fig. 7. T2-weighted (Fig. 7(a)), contrast-enhanced T1-weighted (Fig. 7(f)) and fluid-attenuated inversion recovery (Fig. 7(k)) images are used for tumor diagnosis. The single-shot EPI DWI and the corresponding mean diffusivity (MD) and fractional anisotropy (FA) images are employed as references for evaluation (Fig. 7(b, g, l)), since they do not involve multi-shot imaging thus are free of inter-shot motion artifacts.

The left frontal lobe lesion presents high signal on T2-weighted image (Fig. 7(a)), no obvious enhancement on contrast-enhanced T1-weighted image (Fig. 7(f)), and low signal on fluid-attenuated inversion recovery image (Fig. 7(k)) and single-shot DWI (Figs. 7(b)), indicating that the lesion area is the signal of postoperative fluid accumulation. Thus, there is no fibrous link in the lesion area, and FA should be no signal.

However, the MD and FA of MoDL differs greatly from references (Figs. 7(c, h, n)). PAIR introduces wrong structural features in the reconstructed images (yellow arrow in Fig. 7(d)) because the displayed region is larger than the reference regions in the structured image (Figs. 7(a, f, k) or single-shot DWI image (Fig. 7(b)).

The proposed PIDD provides DWI images without noticeable motion artifacts and incorrect structures (Figs. 7(e, j, o)). PIDD also has the matched lesion region with reference DWI images (Fig. 7(e)). Moreover, the lesion area in FA of PIDD (blue arrow in Fig. 7(l)) shows no signal, suggesting no fibrous link, which is more consistent with the real pathology than other FA images (blue arrows in Fig. 7(m, o)).

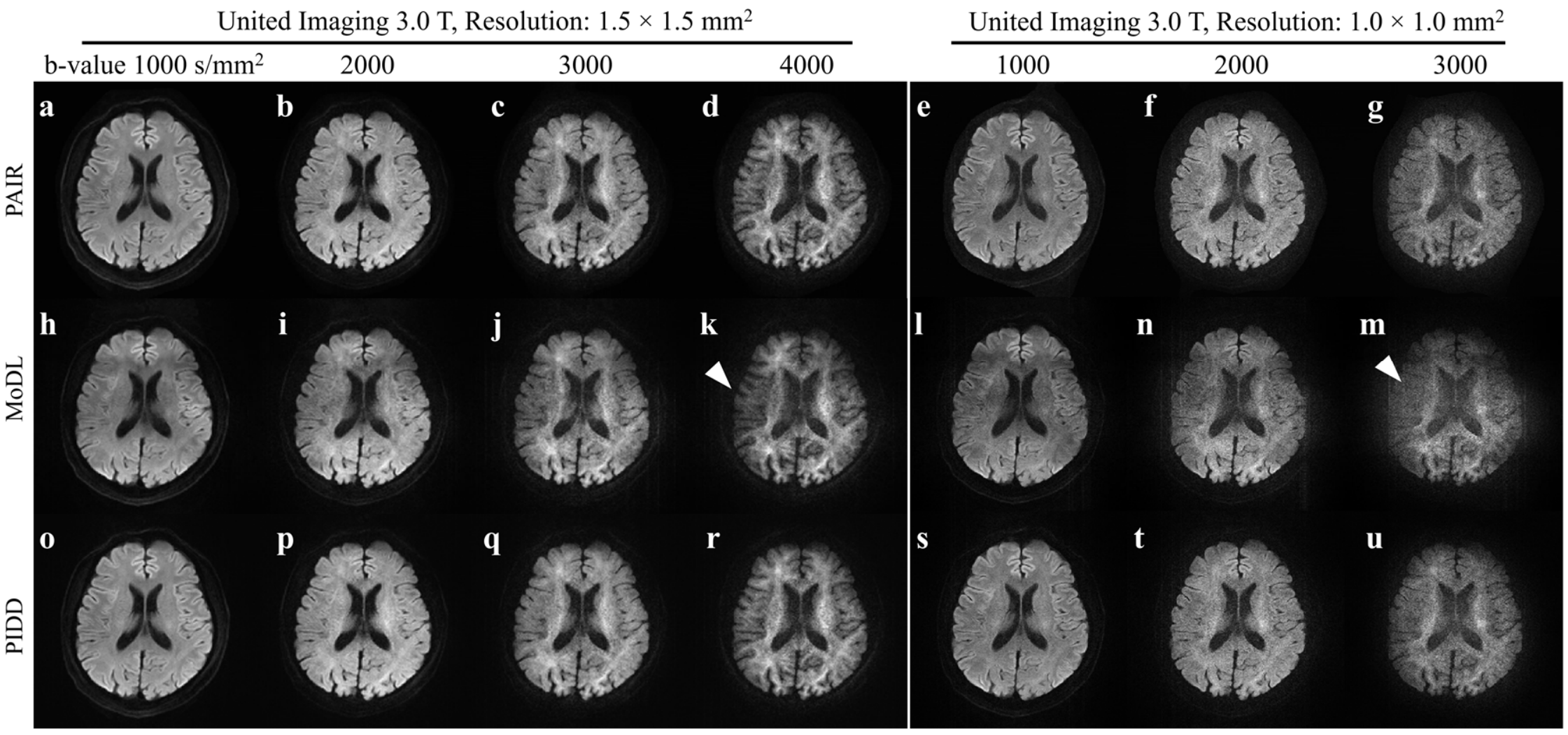

**Fig. 7.** Generalization study on **DATASET II** of healthy volunteer DWI. (a)-(u) are trace DWI images calculated by three diffusion direction DWI images. The incorrect contrast induced by motion artifacts and noise residual are marked by white arrows. Note: Reconstruction parameters of PAIR are optimized on each image to get the best removal of motion artifacts. MoDL is trained on **DATASET I** with 1296 in vivo real samples. PIDD is trained with 100,000 synthetic samples.

From a diagnostic perspective, a reader study (Fig. 7(p)) is conducted to evaluate PIDD on patient data. Seven readers (five radiologists with 5/8/13/20/30 years' experience and two neurosurgeons with 9/26 years' experience) are invited to independently score the image quality of reconstruction. Clinically used low-resolution single-shot DWI images, which are free of inter-shot motion artifacts, are provided as references. Readers are blind to all patient information and reconstruction methods.

In total, 43 slices from 9 patients containing tumors or diseased regions are selected for evaluation, and each slice has 16 diffusion directions. For each slice, reconstructed DWI and generated FA are used for scoring. Three clinical criteria are employed for evaluation: signal-to-noise ratio (SNR), artifact suppression, and overall image quality. Each criterion's score is ranged from 0 to 5 with a precision of 0.1. The difference in scores between PIDD and the compared methods are analyzed with the Wilcoxon signed-rank test.

For FA estimated from 16 diffusion directions, PIDD has achieved scores in the range of 3.85~3.90, which show statistical advantages over PAIR and MoDL ($p$<0.001). For DWI images, scores of PAIR and PIDD are around 3.81, which are similar and show no statistical difference. From a diagnostic perspective, PIDD achieves scores that lie in the range of good (3.0~4.0) and are close to excellence (4.0~5.0), and are significantly superior to MoDL.

Reader study demonstrates that PIDD can provide high-resolution DWIs of patients that have good and even approach excellent diagnostic values.

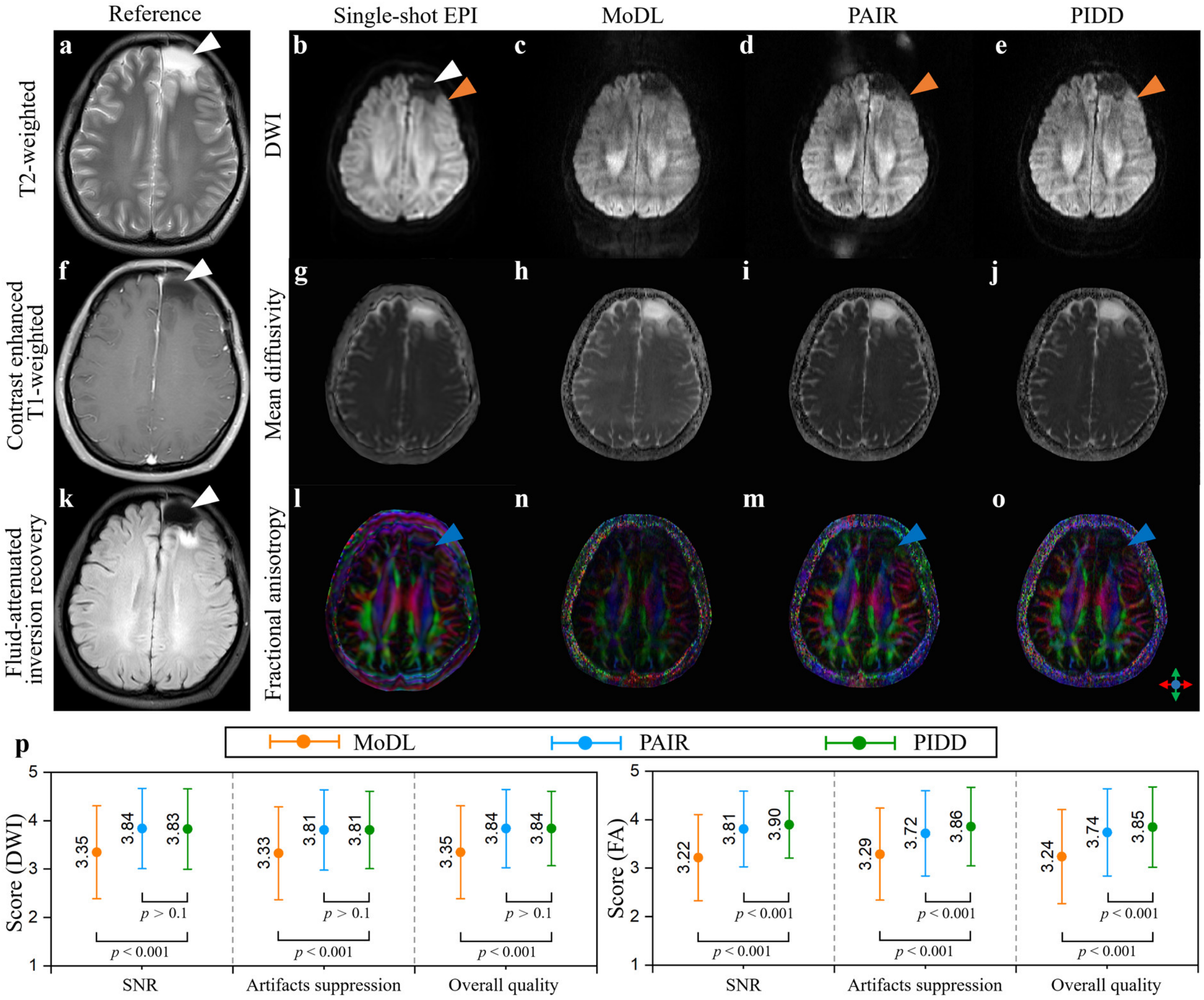

Fig. 8. Clinical study on **DATASET III** patient DTI. (a)-(o) are the reconstructed patient DWI images by MoDL (1.4 seconds per image), PAIR (287.3 seconds per image), and PIDD (2 seconds per image). T2 weighted, contrast enhanced T1 weighted, fluid attenuated inversion recovery, and single-shot EPI DWI images without inter-shot motion artifacts are employed as references, and the region of interest are marked by white arrows. The incorrect structures are marked by yellow arrows. (p) are the score comparisons of the reader study on selected DWI images (43 slices contains tumors or diseased regions from nine patients). Note: The mean values and standard deviations are computed over all tested patients. Wilcoxon signed-rank test are used for scores analysis. $p$<0.001 indicates the difference between two compared methods is statistically significant, while $p$>0.1 indicates no significant difference.

## V. Conclusion

In summary, we propose PIDD, a novel and promising physics-informed deep learning framework, for exploiting the power of diffusion MRI physics to break the bottleneck of obtaining high-quality training data in ms-iEPI DWI reconstruction.

PIDD presents an easy and cost-effective way to synthesize large-scale paired DWI databases for network training. It outperforms state-of-the-art optimization and deep learning methods both in healthy and patient studies, providing more accurate diffusion quantitative maps for patients with seven experienced doctors' verifications, indicating its nice performance in multiple scenarios, and good clinical adaptability.

## Acknowledgments

The authors thank Dr. Mathews Jacob for sharing their codes online, Qiaoli Yao, and Anjie Xie, for assistances in data collection.

## References

[1] F. Knoll *et al*., “Deep-learning methods for parallel magnetic resonance imaging reconstruction: A survey of the current approaches, trends, and issues,” *IEEE Signal Processing Magazine*, vol. 37, no. 1, pp. 128-140, 2020.

[2] O. Perlman *et al*., “Quantitative imaging of apoptosis following oncolytic virotherapy by magnetic resonance fingerprinting aided by deep learning,” *Nature Biomedical Engineering*, vol. 6, no. 5, pp. 648-657, 2022.

[3] V. Vishnevskiy, J. Walheim, and S. Kozerke, “Deep variational network for rapid 4D flow MRI reconstruction,” *Nature Machine Intelligence*, vol. 2, no. 4, pp. 228-235, 2020.

[4] S. Wang *et al*., “Accelerating magnetic resonance imaging via deep learning,” in *2016 IEEE 13th International Symposium on Biomedical Imaging (ISBI)*, pp. 514-517, 2016.

[5] Z. Wang *et al*., “One-dimensional deep low-rank and sparse network for accelerated MRI,” *IEEE Transactions on Medical Imaging*, vol. 42, no. 1, pp. 79-90, 2022.

[6] S. Liu, K.-H. Thung, L. Qu, W. Lin, D. Shen, and P.-T. Yap, “Learning MRI artefact removal with unpaired data,” *Nature Machine Intelligence*, vol. 3, no. 1, pp. 60-67, 2021.

[7] H. Aggarwal, M. Mani, and M. Jacob, “MoDL-MUSSELS: Model-based deep learning for multi-shot sensitivity-encoded diffusion MRI,” *IEEE transactions on Medical Imaging*, vol. 39, no. 4, pp. 1268-1277, 2019.

[8] A. Gao *et al*., “Whole-tumor histogram analysis of multiple diffusion metrics for glioma genotyping,” *Radiology*, vol. 302, no. 3, pp. 652-661, 2022.

[9] J. A. McKay *et al*., “A comparison of methods for high-spatial-resolution diffusion-weighted imaging in breast MRI," *Radiology*, vol. 297, no. 2, pp. 304-312, 2020.

[10] P. Mukherjee, J. Berman, S. Chung, C. Hess, and R. Henry, “Diffusion tensor MR imaging and fiber tractography: Theoretic underpinnings,” *American Journal of Neuroradiology*, vol. 29, no. 4, pp. 632-641, 2008.

[11] H. An, X. Ma, Z. Pan, H. Guo, and E. Y. P. Lee, “Qualitative and quantitative comparison of image quality between single-shot echo-planar and interleaved multi-shot echo-planar diffusion-weighted imaging in female pelvis,” *European Radiology*, vol. 30, no. 4, pp. 1876-1884, 2020.

[12] S. Skare, R. D. Newbould, D. B. Clayton, G. W. Albers, S. Nagle, and R. Bammer, “Clinical multishot DW-EPI through parallel imaging with considerations of susceptibility, motion, and noise,” *Magnetic Resonance in Medicine*, vol. 57, no. 5, pp. 881-890, 2010.

[13] N.-k. Chen, A. Guidon, H.-C. Chang, and A. W. Song, “A robust multi-shot scan strategy for high-resolution diffusion weighted MRI enabled by multiplexed sensitivity-encoding (MUSE),” *NeuroImage*, vol. 72, pp. 41-47, 2013.

[14] A. W. Anderson and J. C. Gore, “Analysis and correction of motion artifacts in diffusion weighted imaging,” *Magnetic Resonance in Medicine*, vol. 32, no. 3, pp. 379-87, 1994.

[15] X. Ma, Z. Zhang, E. Dai, and H. Guo, “Improved multi-shot diffusion imaging using GRAPPA with a compact kernel,” *NeuroImage*, vol. 138, pp. 88-99, 2016.

[16] M. Herbst, J. Maclaren, M. Weigel, J. Korvink, J. Hennig, and M. Zaitsev, “Prospective motion correction with continuous gradient updates in diffusion weighted imaging,” *Magnetic Resonance in Medicine*, vol. 67, no. 2, pp. 326-338, 2012.

[17] Y. Hu *et al*., “RUN-UP: Accelerated multi-shot diffusion-weighted MRI reconstruction using an unrolled network with U-Net as priors,” *Magnetic Resonance in Medicine*, vol.85, no.2, pp. 709-720, 2020.

[18] H. Zhang *et al*., “Deep learning based multiplexed sensitivity-encoding (DL-MUSE) for high-resolution multi-shot DWI,” *NeuroImage*, vol. 244, p. 118632, 2021.

[19] M. Mani, H. K. Aggarwal, V. Magnotta, and M. Jacob, “Improved MUSSELS reconstruction for high-resolution multi-shot diffusion weighted imaging,” *Magnetic Resonance in Medicine*, vol. 83, no. 6, pp. 2253-2263, 2020.

[20] M. Mani, M. Jacob, D. Kelley, and V. Magnotta, “Multi-shot sensitivity-encoded diffusion data recovery using structured low-rank matrix completion (MUSSELS),” *Magnetic resonance in medicine*, vol. 78, no. 2, pp. 494-507, 2017.

[21] C. Qian *et al*., “A paired phase and magnitude reconstruction for advanced diffusion-weighted imaging,” *IEEE Transactions on Biomedical Engineering*, vol. 70, no. 12, pp. 3425-3435, 2023.

[22] G. E. Karniadakis, I. G. Kevrekidis, L. Lu, P. Perdikaris, S. Wang, and L. Yang, “Physics-informed machine learning,” *Nature Reviews Physics*, vol. 3, no. 6, pp. 422-440, 2021.

[23] Q. Yang, Z. Wang, K. Guo, C. Cai, and X. Qu, “Physics-driven synthetic data learning for biomedical magnetic resonance: The imaging physics-based data synthesis paradigm for artificial intelligence,” *IEEE Signal Processing Magazine*, vol. 40, no. 2, pp. 129-140, 2023.

[24] C. Qian *et al*., “Physics-informed deep diffusion MRI reconstruction: Break the bottleneck of Training data in artificial intelligence," in *2023 IEEE 20th International Symposium on Biomedical Imaging (ISBI)*, pp. 1-5, 2023.

[25] M. Uecker *et al*., “ESPIRiT—An eigenvalue approach to autocalibrating parallel MRI: Where SENSE meets GRAPPA,” *Magnetic Resonance in Medicine*, vol. 71, no. 3, pp. 990-1001, 2014.

[26] J. Zbontar *et al*., “fastMRI: An open dataset and benchmarks for accelerated MRI,” arXiv preprint arXiv:1811.08839, 2018.

[27] Q. Tian *et al*., “DeepDTI: High-fidelity six-direction diffusion tensor imaging using deep learning,” *NeuroImage*, vol. 219, p. 117017, 2020.

[28] D. Le Bihan *et al*., “Diffusion tensor imaging: concepts and applications,” *Journal of Magnetic Resonance Imaging*, vol. 13, no. 4, pp. 534-546, 2001.

[29] E. Garyfallidis *et al*., “Dipy, a library for the analysis of diffusion MRI data,” *Frontiers in Neuroinformatics*, vol. 8, p. 8, 2014.

[30] R. Bammer, “Basic principles of diffusion-weighted imaging,” *European Journal of Radiology*, vol. 45, no. 3, pp. 169-184, 2003.

[31] D. Greitz, R. Wirestam, A. Franck, B. Nordell, C. Thomsen, and F. Ståhlberg, “Pulsatile brain movement and associated hydrodynamics studied by magnetic resonance phase imaging: The Monro-Kellie doctrine revisited,” *Neuroradiology*, vol. 34, pp. 370-380, 1992.

[32] J.-F. Cai, E. J. Candès, and Z. Shen, “A singular value thresholding algorithm for matrix completion,” *SIAM Journal on Optimization*, vol. 20, no. 4, pp. 1956-1982, 2010.

[33] J. P. Haldar, “Low-rank modeling of local k -space neighborhoods (LORAKS) for constrained MRI,” *IEEE Transactions on Medical Imaging*, vol. 33, no. 3, pp. 668-681, 2013.

[34] J. A. McKay, S. Moeller, L. Zhang, E. J. Auerbach, M. T. Nelson, and P. J. Bolan, “Nyquist ghost correction of breast diffusion weighted imaging using referenceless methods,” *Magnetic Resonance in Medicine*, vol. 81, no. 4, pp. 2624-2631, 2019.

[35] T. Zhang, J. M. Pauly, S. S. Vasanawala, and M. Lustig, “Coil compression for accelerated imaging with Cartesian sampling,” *Magnetic Resonance in Medicine*, vol. 69, no. 2, pp. 571-582, 2013.

[36] M. Giannelli, S. Diciotti, C. Tessa, and M. Mascalchi, “Characterization of Nyquist ghost in EPI-fMRI acquisition sequences implemented on two clinical 1.5 T MR scanner systems: Effect of readout bandwidth and echo spacing,” *Journal of Applied Clinical Medical Physics*, vol. 11, no. 4, pp. 170-180, 2010.

[37] Y. Zhou *et al*., “CloudBrain-ReconAI: A cloud computing platform for MRI reconstruction and radiologists' image quality evaluation,” *IEEE Transactions on Cloud Computing*, vol. 12, no. 4, pp. 1359-1371, 2024.